\documentclass[dvips]{arxstspdf}
\usepackage{graphics,flushend}
\volume{25}
\issue{2}
\pubyear{2010}
\firstpage{162}
\lastpage{165}
\doi{10.1214/10-STS308B}
\referstodoi{10.1214/09-STS308}

\begin{document}
\begin{frontmatter}

\title{Bayesian Statistics Then and Now\thanksref{TC}}
\runtitle{Comment}

\thankstext{TC}{Discussion of Bradley
Efron's article, ``The future of indirect evidence'' and of Robert
Kass's discussion of Efron's article.}
%To appear in \textit{Statistical Science.}}

\begin{aug}
\author{\fnms{Andrew}
\snm{Gelman}\ead[label=e1]{gelman@stat.columbia.edu}\ead[label=u1,url]{http://www.stat.columbia.edu/\texttildelow gelman}}
\runauthor{A. Gelman}

\affiliation{Columbia University}

\address{Andrew Gelman is Professor,
Department of Statistics and Department of Political Science, Columbia University
 \printead{e1,u1}}

\end{aug}

% ABSTRACT

% KEYWORDS

\end{frontmatter}

It is always a pleasure to hear Brad Efron's thoughts on the next
century of statistics, especially considering the huge influence he has
had on the field's present state and future directions, both in
model-based and nonparametric inference.

\section*{Three Meta-Principles of Statistics}

Before going on, I would like to state three meta-principles of
statistics which I think are relevant to the current discussion.

First, \textit{the information principle}, which is that the key to a
good statistical method is not its underlying philosophy or mathematical
reasoning, but rather what information the method allows us to use. Good
methods make use of more information. This can come in different ways:
in my own experience (following the lead of Efron and Morris, \citeyear{EfronMorris1971},
among others), hierarchical Bayes allows us to combine different data
sources and weight them appropriately using partial pooling. Other
statisticians find parametric Bayes too restrictive: in practice,
parametric modeling typically comes down to conventional models such as
the normal and gamma distributions, and the resulting inference does not
take advantage of distributional information beyond the first two
moments of the data. Such problems motivate more elaborate models, which
raise new concerns about overfitting, and so on.

As in many areas of mathematics, theory and practice leapfrog each
other: as Efron notes, empirical Bayes methods have made great practical
advances but ``have yet to form into a coherent theory.'' In the past
few decades, however, with the work of Lindley and Smith (\citeyear{LindleySmith1972}) and many
others, empirical Bayes has been folded into hierarchical Bayes, which
is part of a coherent theory that includes inference, model checking,
and data collection (at least in my own view, as represented in chapters
6 and 7 of Gelman et al., \citeyear{GelmanEtAl2003}). Other times, theoretical and even
computational advances lead to practical breakthroughs, as Efron
illustrates in his discussion of the progress made in genetic analysis
following the Benjamini and Hochberg paper on false discovery rates.

My second meta-principle of statistics is \textit{the methodological
attribution problem,} which is that the many useful contributions of a
good statistical consultant, or collaborator, will often be attributed
to the statistician's methods or philosophy rather than to the artful
efforts of the statistician himself or herself. Don Rubin has told me
that scientists are fundamentally Bayesian (even if they do not realize
it), in that they interpret uncertainty intervals Bayesianly. Brad Efron
has talked vividly about how his scientific collaborators find
permutation tests and $p$-values to be the most convincing form of
evidence. Judea Pearl assures me that graphical models describe how
people really think about causality. And so on. I am sure that all these
accomplished researchers, and many more, are describing their
experiences accurately. Rubin wielding a posterior distribution is a
powerful thing, as is Efron with a permutation test or Pearl with a
graphical model, and I believe that (a) all three can be helping people
solve real scientific problems, and (b) it is natural for their
collaborators to attribute some of these researchers' creativity to
their methods.

The result is that each of us tends to come away from a collaboration or
consulting experience with the warm feeling that our methods really
work, and that they represent how scientists really think. In stating
this, I am not trying to espouse some sort of empty pluralism---the
claim that, for example, we would be doing just as well if we were all
using fuzzy sets, or correspondence analysis, or some other obscure
statistical method. There is certainly a reason that methodological
advances are made, and this reason is typically that existing methods
have their failings. Nonetheless, I think we all have to be careful
about attributing too much from our collaborators' and clients'
satisfaction with our methods.

My third meta-principle is that \textit{different applications demand
different philosophies.} This principle comes up for me in Efron's
discussion of hypothesis testing and the so-called false discovery rate,
which I label as ``so-called'' for the following reason. In Efron's
formulation (which follows the classical multiple comparisons
literature), a ``false discovery'' is a zero effect that is identified
as nonzero, whereas, in my own work, I never study zero effects. The
effects I study are sometimes small but it would be silly, for example,
to suppose that the difference in voting patterns of men and women
(after controlling for some other variables) could be exactly zero. My
problems with the ``false discovery'' formulation are partly a matter of
taste, I'm sure, but I believe they also arise from the difference
between problems in genetics (in which some genes really have
essentially zero effects on some traits, so that the classical
hypothesis-testing model is plausible) and in social science and
environmental health (where essentially everything is connected to
everything else, and effect sizes follow a continuous distribution
rather than a mix of large effects and near-exact zeroes).

To me, the false discovery rate is the latest flavor-of-the-month
attempt to make the Bayesian omelette without breaking the eggs. As
such, it can work fine if the implicit prior is ok, it can be a great
method, but I really don't like it as an underlying principle, as it is
all formally based on a hypothesis-testing framework that, to me, is
more trouble than it's worth. In thinking about multiple comparisons in
my own research, I prefer to discuss errors of Type S and Type M rather
than Type~1 and Type 2 (Gelman and Tuerlinckx, \citeyear{GelmanTuerlinckx2000};
Gelman and Weakliem, \citeyear{GelmanWeakliem2009};
Gelman, Hill and Yajima, \citeyear{GelmanHillYajima2009}). My point here, though,
is simply that any given statistical concept will make more sense in
some settings than others.

For another example of how different areas of application merit
different sorts of statistical thinking, consider Rob Kass's remark: ``I
tell my students in neurobiology that in claiming statistical
significance I get nervous unless the $p$-value is much smaller than
0.01.'' In political science, we are typically not aiming for that level
of uncertainty. (Just to get a sense of the scale of things, there have
been barely 100 national elections in all of U.S. history, and political
scientists studying the modern era typically start in 1946.)

\section*{Progress in Parametric Bayesian Inference}

\begin{figure*}[t]

\includegraphics{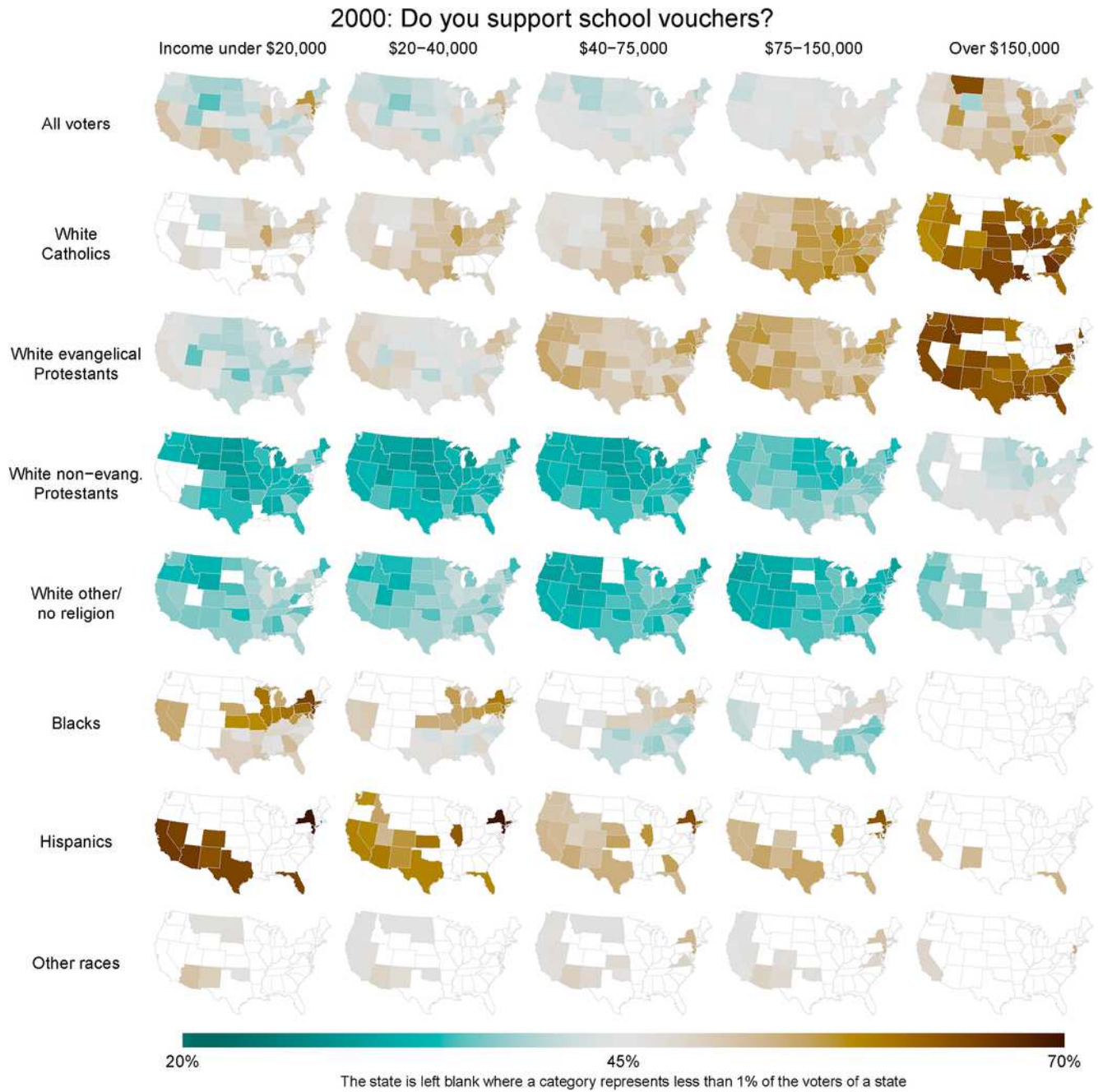}

\caption{Estimated proportion of voters in each state who support
federal spending on school vouchers, broken down by religion/ethnicity
and income categories. The estimates come from a hierarchical Bayesian
analysis fit to data from the National Annenberg Election Survey,
adjusted to population and voter turnout data from the U.S.
Census.}\label{fig1}
\end{figure*}

I also think that Efron is doing parametric Bayesian inference a
disservice by focusing on a fun little baseball example that he and
Morris worked on 35 years ago. If he would look at what is being done
now, he would see all the good statistical practice that, in his section
10, he naively (I think) attributes to ``frequentism.'' Figure~\ref{fig1}
illustrates with a grid of maps of public opinion by state, estimated
from national survey data. Fitting this model took a lot of effort which
was made possible by working within a hierarchical regression
framework---``a good set of work rules,'' to use Efron's expression.
Similar models have been used recently to study opinion trends in other
areas such as gay rights in which policy is made at the state level, and
so we want to understand opinions by state as well (Lax and Phillips, \citeyear{LaxPhillips2009}).

%f1 ###

I also completely disagree with Efron's claim that frequentism (whatever
that is) is ``fundamentally conservative.'' One thing that
``frequentism'' absolutely encourages is for people to use horrible,
noisy estimates out of a fear of ``bias.'' More generally, as discussed
by Gelman and Jakulin (\citeyear{GelmanJakulin2007}), Bayesian inference is conservative in that
it goes with what is already known, unless the new data force a change.
In contrast, unbiased estimates and other unregularized classical
procedures are noisy and get jerked around by whatever data happen to
come by---not really a conservative thing at all. To make this argument
more formal, consider the multiple comparisons problem. Classical
unbiased comparisons are noisy and must be adjusted to avoid
overinterpretation; in contrast, hierarchical Bayes estimates of
comparisons are conservative (when two parameters are pulled toward a
common mean, their difference is pulled toward zero) and less likely to
appear to be statistically significant (Gelman and Tuerlinckx, \citeyear{GelmanTuerlinckx2000}).

Another way to understand this is to consider the ``machine learning''
problem of estimating the probability of an event on which we have very
little direct data. The most conservative stance is to assign a
probability of $\tfrac12$; the next-conservative approach might be to
use some highly smoothed estimate based on averaging a large amount of
data; and the unbiased estimate based on the local data is hardly
conservative at all! Figure~\ref{fig1} illustrates our conservative estimate of
public opinion on school vouchers. We prefer this to a noisy,
implausible map of unbiased estimators.

Of course, frequentism is a big tent and can be interpreted to include
all sorts of estimates, up to and including whatever Bayesian thing I
happen to be doing this week---to make any estimate ``frequentist,'' one
just needs to do whatever combination of theory and simulation is
necessary to get a sense of my method's performance under repeated
sampling. So maybe Efron and I are in agreement in practice, that any
method is worth considering if it works, but it might take some work to
see if something really does indeed work.

\section*{Comments on Kass's Comments}

Before writing this discussion, I also had the opportunity to read Rob
Kass's comments on Efron's article.

I pretty much agree with Kass's points, except for his claim that most
of Bayes is essentially maximum likelihood estimation. Multilevel
modeling is only approximately maximum likelihood if you follow Efron
and Morris's empirical Bayesian formulation in which you average over
intermediate parameters and maximize over hyperparameters, as I gather
Kass has in mind. But then this makes ``maximum likelihood'' a matter of
judgment: what exactly is a hyperparameter? Things get tricky with
mixture models and the like. I guess what I'm saying is that maximum
likelihood, like many classical methods, works pretty well in practice
only because practitioners interpret the methods flexibly and do not do
the really stupid versions (such as joint maximization of parameters and
hyperparameters) that are allowed by the theory.

Regarding the difficulties of combining evidence across species (in
Kass's discussion of the DuMouchel and Harris paper), one point here is
that this works best when the parameters have a real-world meaning. This
is a point that became clear to me in my work in toxicology (Gelman,
Bois and Jiang, \citeyear{GelmanBoisJiang1996}): when you have a model whose parameters have
numerical interpretations (``mean,'' ``scale,''\break ``curvature,'' and so
forth), it can be hard to get useful priors for them, but when the
parameters have substantive interpretations (``blood flow,''
``equilibrium concentration,'' etc.), then this opens the door for real
prior information. And, in a hierarchical context, ``real prior
information'' does not have to mean a specific, pre-assigned prior;
rather, it can refer to a model in which the parameters have a
group-level distribution. The more real-worldy the parameters are, the
more likely this group-level distribution can be modeled accurately. And
the smaller the group-level error, the more partial pooling you will get
and the more effective your Bayesian inference is. To me, this is the
real connection between scientific modeling and the mechanics of
Bayesian smoothing, and Kass alludes to some of this in the final
paragraph of his comment.

Hal Stern once said that the big divide in statistics is not between
Bayesians and non-Bayesians but rather between modelers and
non-modelers. And, indeed, in many of my Bayesian applications, the big
benefit has come from the likelihood. But sometimes that is because we
are careful in deciding what part of the model is ``the likelihood.''
Nowadays, this is starting to have real practical consequences even in
Bayesian inference, with methods such as DIC, Bayes factors, and
posterior predictive checks, all of whose definitions depend crucially
on how the model is partitioned into likelihood, prior, and hyperprior
distributions.

On one hand, I am impressed by modern machine-learning methods that
process huge datasets with I agree with Kass's concluding remarks that
emphasize how important it can be that the statistical methods be
connected with minimal assumptions; on the other hand, I appreciate
Kass's concluding point that statistical methods are most powerful when
they are connected to the particular substantive question being studied.
I agree that statistical theory is far from settled, and I agree with
Kass that developments in Bayesian modeling are a promising way to move
forward.

\section*{Acknowledgments}

We thank Yu-Sung Su and Daniel Lee for collaboration on Figure
1, Rob Kass for helpful comments, and the NSF, %National Science Foundation,
National Institutes of Health, Institute of Educational Sciences,
National Security Agency, Department of Energy, and the Columbia
University Applied Statistics Center for partial support of this work.
The research reported here was supported by the Institute of Education
Sciences, US Department of Education, through Grant R305D090006 to
Columbia University.


\begin{thebibliography}{99}

%b1 ###
\bibitem[\protect\citeauthoryear{}{1971}]{EfronMorris1971}
\textsc{Efron, B.} and \textsc{Morris, C.} (1971).
Limiting the risk of Bayes and empirical Bayes estimates---Part I: The Bayes case.
\textit{J.~Amer. Statist. Assoc.} \textbf{66} 807--815.
\MR{0323014}

%b2 ###
\bibitem[\protect\citeauthoryear{}{1996}]{GelmanBoisJiang1996}
\textsc{Gelman, A., Bois, F. Y.} and \textsc{Jiang, J.} (1996).
Physiological pharmacokinetic analysis using population modeling and informative prior
distributions.
\textit{J.~Amer. Statist. Assoc.} \textbf{91} 1400--1412.


%b4 ###
\bibitem[\protect\citeauthoryear{}{2009}]{GelmanHillYajima2009}
\textsc{Gelman, A., Hill, J.} and \textsc{Yajima, M.} (2009).
Why we (usually) don't have to worry about multiple comparisons. Technical report, Dept.
Statistics, Columbia Univ.


%b7 ###
\bibitem[\protect\citeauthoryear{}{2007}]{GelmanJakulin2007}
\textsc{Gelman, A.} and \textsc{Jakulin, A.} (2007).
Bayes: Radical, liberal, or conservative?
\textit{Statist. Sinica} \textbf{17} 422--426.
\MR{2435285}

%b5 ###
\bibitem[\protect\citeauthoryear{}{2000}]{GelmanTuerlinckx2000}
\textsc{Gelman, A.} and \textsc{Tuerlinckx, F.} (2000).
Type S error rates for classical
and Bayesian single and multiple comparison procedures.
\textit{Comput. Statist.} \textbf{15} 373--390.

%b6 ###
\bibitem[\protect\citeauthoryear{}{2009}]{GelmanWeakliem2009}
\textsc{Gelman, A.} and \textsc{Weakliem, D.} (2009).
Of beauty, sex, and power:
Statistical challenges in estimating small effects.
\textit{Amer. Sci.} \textbf{97} 310--316.




%b3 ###
\bibitem[\protect\citeauthoryear{}{2003}]{GelmanEtAl2003}
\textsc{Gelman, A., Carlin, J. B., Stern, H. S.} and \textsc{Rubin, D. B.} (2003).
\textit{Bayesian Data Analysis}, 2nd ed. Chapman and Hall, London.
\MR{2027492}

%b8 ###
\bibitem[\protect\citeauthoryear{}{2009}]{LaxPhillips2009}
\textsc{Lax, J. R.} and \textsc{Phillips, J. H.} (2009).
Gay rights in the states: Public
opinion and policy responsiveness.
\textit{American Political Science Review} \textbf{103}.

%b9 ###
\bibitem[\protect\citeauthoryear{}{1972}]{LindleySmith1972}
\textsc{Lindley, D. V.} and \textsc{Smith, A. F. M.} (1972).
Bayes estimates for the
linear model.
\textit{J.~Roy. Statist. Soc. Ser. B} \textbf{34} 1--41.
\MR{0415861}

\end{thebibliography}
\end{document}